# Warum Astrologie nicht funktionieren kann

Florian Freistetter
Astronomisches Recheninstitut
Universität Heidelberg

Wie untersucht man Astrologie am besten? Man kann ihre Entwicklung aus historischer Sicht betrachten. Man kann die Aussagen der Astrologie aus soziologischer, linguistischer oder statistischer Sicht analysieren und so herausfinden ob sie das leistet, was sie verspricht, nämlich verlässliche, reproduzierbare und nachprüfbare Informationen. Man kann die Grundlagen der Astrologie aus naturwissenschaftlicher bzw. astronomischer Perspektive angehen, um festzustellen, ob sie eine logische Grundlage hat oder nicht.

Zweierlei ist einer derartigen Analyse vorwegzuschicken: Astrologie existiert seit langer Zeit, ist ein sehr komplexes Phänomen und kann daher in einem so kurzen Text auch nur unvollständig betrachtet werden. Weiterhin ist die Frage nach der Berechtigung einer kritischen Hinterfragung der Astrologie zu stellen. Ein Vorwurf, den Astrologen gegenüber Kritikern sehr oft äußern ist folgender: Gerade weil die Astrologie so komplex ist, darf sie von jemandem, der kein Astrologe ist, nicht kritisiert werden. Man müsse sich erst jahrelang mit der Astrologie beschäftigen, sonst sei man als unwissender Laie nicht qualifiziert zu beurteilen, ob Astrologie funktioniert oder nicht. Selbstverständlich sollte man ein wenig Ahnung von dem haben, was man kritisieren möchte. Aber man muss nicht jahrelang als Astrologe gearbeitet haben, um feststellen zu können, ob sie funktionieren kann oder nicht. Man muss ja auch kein KFZ-Mechaniker sein, um zu überprüfen, ob das Auto, das man kaputt in die Werkstatt gebracht hat, danach wieder fahrtüchtig ist. Wiederum gestehen Astrologen ihrem Publikum - das im Allgemeinen keine professionellen Astrologen umfasst - zu, sich ein Urteil über ihre Arbeit bilden. Demgemäß kann man auch an dieser Stelle die Astrologie untersuchen und sogar kritisieren, auch wenn man selbst kein Astrologe ist.

Außerdem wird sich dieser Text ganz bewusst mit der Astrologie auf einem sehr allgemeinem Niveau beschäftigen, für das keine astrologischen Spezialkenntnisse nötig sind. Denn *die* Astrologie gibt es ohnehin nicht. Wenn man etwa die Horoskope in den Zeitungen und Magazinen kritisiert, dann würden die Astrologen sofort dagegen halten, dass es sich dabei um keine echte Astrologie handle. So haben sich im Dezember 2010 die in der österreichischen Wirtschaftskammer organisierten Astrologen von der astrologischen Prophezeiung konkreter Ereignisse distanziert und diese als „unseriös"[1] bezeichnet. Würde man etwa die Astrologie der Hamburger Schule[2] in Frage stellen, würden sich die Astrologen anderer Schulen nicht angesprochen fühlen. Die Astrologie ist in sich so zersplittert - und das gilt sowohl für die westlich-abendländische Astrologie als auch für die vielen anderen Systeme, so aus Indien oder China - dass die Kritik an einer einzigen Richtung kaum sinnvoll ist. Astrologen können diese Kritik dann immer mit der Variation eines logischen

---

[1] Vgl. http://www.ots.at/presseaussendung/OTS_20101227_OTS0045/gewerbliche-astrologen-distanzieren-sich-von-prognosen-konkreter-ereignisse [Stand: 09.03.2011].

[2] Als Hamburger Schule wird eine bestimmte astrologische Auswertungsmethode bezeichnet, die Anfang des 20. Jahrhunderts von Alfred Witten begründet wurde. Neben dem Postulat eines speziellen Häusersystems zeichnet sich die Hamburger Schule vor allem durch die Verwendung der sogenannten „Transneptune" aus. Dabei handelt es sich um acht zusätzliche Himmelskörper die im Horoskop verwendet werden – die aber in der Realität nicht existieren.

Fehlschlusses vom Typ „No True Scotsman"[3] beantworten: „Was kritisiert wird, ist nicht die wahre Astrologie". Jeder Astrologe macht sich im Prinzip seine eigene Astrologie.

Das ist dann auch schon das erste, was einen bei der Frage nach der Aussagefähigkeit der Astrologie stutzig machen könnte: Wenn die Astrologie so extrem uneinheitlich ist, wie kann man dann überhaupt zu verlässlichem Wissen gelangen? Genau an diesem Punkt will die Erklärung ansetzen: Die Astrologie ist nicht nur in ihrer aktuellen Ausprägung gänzlich heterogen. Schon ihre grundlegenden Annahmen und Aussagen, die allen astronomischen Richtungen gemeinsam sind, sind in sich völlig inkonsistent.

Grundlage jeder Astrologie ist die Behauptung, es gebe einen irgendwie gearteten Zusammenhang zwischen den Objekten am Himmel und dem Leben oder dem Schicksal der Menschen. Damit ist nicht unbedingt eine konkrete Kraft gemeint, die Sterne oder Planeten auf Menschen ausüben. Manche Astrologen vertreten zwar auch diese Meinung[4] aber da sich relativ leicht beweisen lässt, dass keine der bekannten Kräfte – und auch keine unbekannte Kraft - in der Lage wäre, das von der Astrologie Verlangte zu leisten, ziehen sich viele Astrologen auf den Standpunkt zurück, dass die Dinge am Himmel nur eine Art Metapher für die Vorgänge am Erdboden sind: „Wie oben, so auch unten". Zwischen Himmel und Erde herrsche eine gewisse *Synchronizität,* weswegen man das Schicksal der Menschen auch an den Sternen und Planeten ablesen könne.

Auch wenn die Details sich extrem unterscheiden können, so ist dieser Zusammenhang zwischen Menschen und Himmelsobjekten doch diejenige Eigenschaft, die einer Astrologie immer zugrunde liegt und auf die sich dieser Text konzentrieren will. Heute kennt die Astronomie unzählige Himmelskkörper. Unser Sonnensystem besteht aus einem Stern, acht Planeten, fünf Zwergplaneten, 168 Monden, Millionen von Asteroiden und Kometen und jeder Menge kleinerer Objekte wie interplanetarer Staub und interplanetares Gas. Außerhalb unseres Sonnensystems gibt es noch Milliarden anderer Sterne unserer Milchstrassengalaxie, von denen ebenfalls viele Planeten haben - über 500 davon sind mittlerweile bekannt -, und Milliarden anderer Galaxien im Universum. Wenn die Astrologie nun tatsächlich eine Lehre oder sogar eine Wissenschaft ist, die verlässlich und reproduzierbar konkrete Information liefert, dann muss auch ihre Grundlage eine gewisse Konsistenz aufweisen. Wenn die Astrologie keine Lehre der absoluten Beliebigkeit ist, dann muss es möglich sein, zu begründen, welche Himmelskörper man bei der Untersuchung des menschlichen Schicksals berücksichtigen muss und warum.

In der Wissenschaft ist das selbstverständlich. Will man mit der wissenschaftlichen Erforschung der Bewegung der Himmelskörper, der „Himmelsmechanik", zum Beispiel herausfinden, ob ein Asteroid mit der Erde kollidieren wird, dann muss man dessen zukünftige Bewegung berechnen. Diese Bewegung hängt von der Gravitationskraft ab, die auf den Asteroiden wirkt, in diesem Fall zuerst und hauptsächlich einmal die der Sonne; sie hat schließlich die meiste Masse in unserem Sonnensystem. Aber auch die anderen Planeten beeinflussen mit ihrer Masse die

---

Bahn des Asteroiden. Sogar die größeren Monde und Asteroiden können einen relevanten gravitativen Einfluss ausüben. Da die Reichweite der Gravitation unendlich ist - sie wird mit der Entfernung zwar immer schwächer aber nie komplett null -, müsste man genau genommen jedes Objekt im Universum und dessen Gravitationskraft berücksichtigen. Aber natürlich ist der Einfluss, etwa des Jupiter, auf einen Asteroiden im Sonnensystem viel größer als der eines Sterns in der 2,5 Milliarden Lichtjahre entfernten Andromedagalaxie. Wenn man Asteroiden in der Nähe der Erde betrachtet, dann kann man sogar schon den Einfluss der Planeten Uranus und Neptun weitestgehend vernachlässigen. Man kann die Bewegung des Asteroiden mit und ohne die von ihnen ausgeübte Gravitationskraft berechnen und sieht dann, dass dies keinen Unterschied ergiebt. Ihr Einfluss spielt für dieses Problem keine Rolle. Und so kann man für jedes himmelsmechanische Problem angeben, welche Himmelskörper Einflussfaktoren sind und welche nicht. Will man die Bewegung der Marsmonde untersuchen, kann man etwa den Zwergplanet Pluto ignorieren. Will man mehr über die Bewegung der Asteroiden im Kuipergürtel wissen, die sich in der Nähe des Pluto befinden, dann muss man Pluto berücksichtigen, um verlässliche Ergebnisse zu erhalten - braucht sich aber nicht um Mars zu kümmern. Man kann also immer präzisieren, welche Himmelskörper aus welchen Gründen eine Rolle spielen. Und wenn man entsprechende wissenschaftliche Ergebnisse veröffentlicht, muss man ebenso darauf verweisen. Jeder Gutachter würde sonst umgehend und zu Recht bemängeln, dass der Autor vergessen hat, die Wahl seines Modells und seiner Parameter - in diesem Fall eben die Wahl der berücksichtigten Himmelskörper - zu rechtfertigen.

Aber wie ist das nun in der Astrologie? Von den acht Planeten, fünf Zwergplaneten, 168 Monden und dem einen Stern werden in der Astrologie normalerweise sieben Planeten (Merkur, Venus, Mars, Jupiter, Saturn, Uranus, Neptun), ein Zwergplanet (Pluto), ein Mond (der Erdmond) und ein Stern (die Sonne) berücksichtigt. Dabei werden Uranus, Neptun und Pluto erst seit dem Zeitpunkt ihrer jeweiligen Entdeckung benutzt; davor ist man anscheinend immer gut ohne sie ausgekommen.Hier stellt sich nun die Frage, wie Astrologen die Wahl dieser Parameter rechtfertigen. Nach welchen Regeln wählen sie diese Himmelskörper aus und verwerfen andere? Es gibt übrigens Astrologen, die die Auswahl erweitert haben und auch einige größere Asteroiden berücksichtigen. Aber auch dann bleibt die Frage nach den Auswahlkriterien bestehen. Wenn die Astrologie keine Lehre der völligen Beliebigkeit ist, dann müsste sie eigentlich eine in sich konsistente Basis haben, mit der sich diese Fragen beantworten lassen.

Klar ist, dass es keine Frage der astronomischen Klassifikation ist. Die Astrologie verwendet zwar immer alle Planeten; vor 1781 vermisste jedoch kein Astrologe den Uranus, vor 1846 vermutete kein Astrologe die Existenz des Neptun (die Astronomen dagegen schon; sie hatten seine Existenz aus der Beobachtung von Bahnstörungen des Uranus vorhergesagt) und auch Pluto fehlte vor 1930 nicht. Selbst heute werden in den Horoskopen nur die acht Planeten in unserem Sonnensystem verwendet und nicht auch die extrasolaren Planeten, obwohl die Bedeutung der Himmelskörper gemäß astrologischer Logik eigentlich nicht entfernungsabhängig sein kann. Denn der weit von der Erde entfernte Pluto ist im Horoskop genauso wichtig wie die nahe Venus. Bei den Zwergplaneten wird überhaupt nur Pluto benutzt und mitunter der größte Hauptgürtelasteroid und kleinste Zwergplanet Ceres. Die restlichen Zwergplaneten - Eris, Haumea und Makemake - kommen in normalen Horoskopen so gut wie nie vor. Noch deutlicher ist es bei den Monden. In den Horoskopen wird generell nur der Erdmond im Horoskop in betracht gezogen, die restlichen 167 Monde - davon sind Io, Ganymed, Callisto und Titan sogar größer als der Erdmond - werden ignoriert. Gleiches gilt für die Asteroiden: manche Astrologen verwenden bisweilen Asteroiden wie Ceres, Pallas, Vesta oder Chiron. Diese unterscheiden sich jedoch durch nichts von den Millionen anderen Asteroiden im Sonnensystem, die nicht in den Horoskopen auftauchen.

Auch physikalische Parameter sind bei der Auswahl der astrologisch relevanten Himmelskörper entscheidend. Der Einfluss des schweren und großen Jupiters im Horoskop ist genauso wichtig wie der des kleinen und leichten Monds oder des Pluto. Die gewaltige Sonne ist ebenso wichtig wie der winzige Mars. Wie der sonnennächste Planet Merkur muss auch der sonnenfernste Planet Neptun im Horoskop berücksichtigt werden. Weiterhin kann die Sichtbarkeit mit bloßem Auge kein astrologisches Auswahlkriterium sein. Denn man kann zwar Sonne, Mond, Merkur, Venus, Mars, Jupiter und Saturn leicht sehen. Aber bei Uranus besteht die freiäugige Sichtbarkeit nur noch theoretisch, Neptun und Pluto sind ohne technische Hilfsmittel unsichtbar.

Viele Astrologen können auf die Frage nach der Auswahl der relevanten Himmelskörper keine Antwort geben. Aber einige, die es doch getan haben,[5] haben meistens und verständlicherweise nicht physikalisch-astronomisch argumentiert, sondern eher emotional: die im Horoskop verwendeten Himmelskörper seien eben genau diejenigen, die für die Menschen auf irgendeine Art und Weise von besonderer Bedeutung wären. Das ist zweifelsohne eine äußerst schwammige Begründung - die noch dazu nicht frei von Widersprüchen ist. Pluto ist beispielsweise heute Teil von nahezu jeder astrologischen Analyse. Dass er mittlerweile nur noch als Zwergplanet klassifiziert wird und nicht mehr als Planet, spielt dabei keine Rolle. Pluto sei in der Astrologie deswegen von Bedeutung, weil er eben lange Zeit als Planet des Sonnensystems geführt wurde und deswegen Bedeutung für die Menschen hat und hatte.[6] Aber folgt man dieser Argumentation, dann müssten der Zwergplanet Ceres und die Asteroiden Pallas, Juno, Vesta und Hygiea ebenso wichtig sein wie Pluto. Denn bei ihrer Entdeckung Anfang des 19. Jahrhunderts wurden diese Himmelskörper ebenfalls als Planeten klassifiziert und erst Jahrzehnte später wieder "degradiert" - so wie Pluto. Da sie die ersten jemals entdeckten Asteroiden waren, hatten sie durchaus große Bedeutung für die Menschen. Ceres war vormals mindestens so bekannt wie Pluto heute. Der Mond der Erde wird laut manchen Astrologen deswegen als einziger der 168 bekannten Monde berücksichtigt, weil er eben der Mond der Erde und deswegen relevant ist. Nicht relevant ist aber der Asteroid 2002 AA99 obwohl er ebenfalls ein koorbitales Objekt der Erde ist (sich also wie der Mond ständig in ihrer Nähe aufhält). Aber auch die vier galileischen Monde des Jupiter - von denen drei größer als der Erdmond sind - haben in der Geschichte der Menschheit eine wichtige Rolle gespielt. Sie waren nämlich die ersten neuen Himmelskörper, die Galileo Galilei 1609 mit seinem Teleskop entdeckte, und ihre Beobachtung beeinflusste den Wandel zum heliozentrischen Weltbild.[7] Warum tauchen diese galileischen Monde jedoch nie in Horoskopen auf? Und was ist mit der Vielzahl an Sternen am Himmel, die für die Menschen ebenso relevant wie die Planeten waren? Sie dienen in der Astrologie nur als Hintergrund, vor dem sich alles abspielt. Egal wie man es betrachtet, es gibt

---

keine objektiven Regeln, die klar darlegen, welche Himmelskörper für die Astrologie relevant sind und welche nicht.

Verständlich wird die Auswahl der Himmelskörper in der Astrologie freilich aus der historischen Entwicklung heraus. Sterne wurden in der Antike nicht als eigenständige Himmelskörper angesehen, sondern nur als Lichtpunkte oder Löcher in der äußersten Himmelssphäre. Von den übrigen Himmelskörpern waren damals nur solche bekannt, die mit direkt sichtbar waren und von denen man dachte, dass sie die Erde als Zentrum umkreisen. Das waren Sonne, Mond, Merkur, Venus, Mars, Jupiter und Saturn. Lange Zeit beschränkte sich die Astrologie daher auf diese Gruppe, bis zwischen 1781 und 1930 die neu entdeckten Planeten Uranus, Neptun und Pluto hinzu kamen, deren Existenz kein einziger Astrologe vorhergesagt hatte. Mittlerweile kennen wir eine Unmenge von verschiedenen Himmelskörpern. Im Gegensatz zum Weltbild der Antike, das immer noch die Grundlage der Astrologie bildet, wissen wir heute, dass die Sterne keine Punkte an einer Himmelssphäre sind, sondern weit entfernte Sonnen. Wir wissen, dass weder die Erde noch die anderen Planeten unseres Sonnensystems etwas Besonders darstellen, sondern dass es im Universum noch unzählige andere Planeten gibt. Wir wissen, dass neben den mit bloßem Auge sichtbaren noch unzählige Himmelskörper existieren, die nur mit technischen Hilfsmitteln beobachtbar sind.

Die Astrologie kann nun natürlich aus rein praktischen Gründen nicht alle Planeten, Zwergplaneten, Monde oder Asteroiden berücksichtigen. Da sie aber auch keine konkreten Regeln zur Auswahl anbieten kann, welche Objekte wann zu verwenden sind und wann nicht, erweist sich diese Lehre im Prinzip als arbiträr. Astrologie kann nicht funktionieren. Ein Astrologe kann sich bei seiner Arbeit entweder an die klassische Astrologie der Antike halten; er kann sich neuerer Himmelskörper bedienen oder sogar – wie in der Hamburger Schule – Himmelskörper verwenden, die überhaupt nicht existieren. Die Astrologie ist ein Wahrsage- und Assoziationssystem, das gänzlich von der physikalischen Realität abgekoppelt ist und dessen Elemente und Auslegungsregeln ausschließlich von den Vorlieben des jeweiligen Astrologen abhängig sind.

Gerade weil die Aussagen und *Analysen* der Astrologen diese Beliebigkeit widerspiegeln, funktioniert die Astrologie anscheinend für viele Leute: Mit einer Lehre, die mangels konsistenter Basis nichts erklären kann, kann man natürlich alles erklären. Je unspezifischer die theoretische Basis, desto unspezifischer sind auch die daraus abgeleiteten Aussagen und desto freier können diese vom Astrologen interpretiert werden. Wer astrologische Texte objektiv betrachtet, stellt fest, dass sie meist äußerst vage sind und als „Barnum-Texte"[8] im Prinzip fast beliebige Deutungen zulassen. Das macht es für den Rezipienten äußerst leicht, diese Aussagen mit seiner Selbstwahrnehmung in scheinbare Übereinstimmung zu bringen und so zur Überzeugung zu kommen, das Horoskop würde eine tatsächliche Beschreibung der Persönlichkeit liefern.[9] Zusammenfassend bleibt festzustellen, dass aus logisch-naturwissenschaftlicher Sicht die Astrologie - egal welcher konkreten Ausprägung - über keine in sich schlüssige Grundlage verfügt, anhand der sich allgemeingültige und überprüfbare Regeln für die astrologische Arbeit ableiten lassen. Die astrologische Deutung bleibt somit beliebig; Astrologie funktioniert nicht.

---

[8] Der „Barnum-Effekt" geht auf den Psychologen Bertram Forer zurück, der die menschliche Tendenz beschrieb, unspezifische und allgemein gehaltene Aussagen über die eigene Persönlichkeit als zutreffend anzusehen. Solche „Barnum-Texte" findet man unter anderem in Horoskopen.

[9] Etwa Charles Richard Snyder: Why Horoscopes Are True. The Effects of Specificity on Acceptance of Astrological interpretations, in: Journal of Clinical Psychology 30 (1974), H. 4, S. 577–580; Bertram Forer: The Fallacy of Personal Validation. A Classroom Study of Gullibility, in Journal of Abnormal and Social Psychology 44 (1949), S. 118-123. Die psychologische oder linguistische Analyse astrologischer Prognosen ist allerdings ein anderes Thema, für dessen detaillierte Untersuchung hier kein Platz bleibt.